# Large-area, high-resolution characterisation and classification of damage mechanisms in dual-phase steel using deep learning

Carl Kusche[1,+], Tom Reclik[1,+], Martina Freund[1], Talal Al-Samman[1], Ulrich Kerzel[1], Sandra Korte-Kerzel[1,*]

[1] Institute of Physical Metallurgy and Metal Physics, RWTH Aachen University, Aachen, Germany

[+] equal contribution

[*] korte-kerzel@imm.rwth-aachen.de

## Abstract

High performance materials, from natural bone over ancient damascene steel to modern superalloys, typically possess a complex structure at the microscale. Their properties exceed those of the individual components and their knowledge-based improvement therefore requires understanding beyond that of the components' individual behaviour. Electron microscopy has been instrumental in unravelling the most important mechanisms of co-deformation and in-situ deformation experiments have emerged as a popular and accessible technique. However, a challenge remains: to achieve high spatial resolution and statistical relevance in combination. Here, we overcome this limitation by using panoramic imaging and machine learning to study damage in a dual-phase steel. This high-throughput approach now gives us strain and microstructure dependent insights into the prevalent damage mechanisms and allows the separation of inclusions and deformation–induced damage across a large area of this heterogeneous material. Aiming for the first time at automated classification of the majority of damage sites rather than only the most distinct, the new method also encourages us to expand current research past interpretation of exemplary cases of distinct damage sites towards the less clear-cut reality.

## Introduction

The damage encountered in forming of metallic components poses both a great challenge in manufacturing, as well as enormous potential for improved lightweight design. Current efforts to lighten load-bearing structures, e.g. car bodies, focus on an increased use of low density materials, such as aluminium and magnesium alloys, enhancement of the (cold) formability of high strength steels and minimisation of the material used in structural elements.

In the context of increasing formability and improving the (remaining) local strength after processing, the key challenge is to comprehend the microstructural damage induced during deformation in metal forming. This would allow engineers to minimise damage by adapting the forming processes and accurately assess the remaining material performance in service after processing to save excess material. However, the physical mechanisms underlying damage in the most prominent high performance materials, such as dual-phase steels, are incompletely understood in spite of an ever-increasing research effort over the last decade [1].

The reason why the fundamental mechanisms of damage formation remain elusive lies in the complex microstructure of dual-phase steels. They are categorized and marketed based only on attained yield strength and the presence of a dual phase (ferrite and martensite) microstructure, but their properties are determined by a zoo of often interdependent parameters. These include volume fraction, morphology and carbon content of the hard martensite phase, mechanical contrast between the phases, homogeneity of texture, grain size and phase distribution as well as local and mesoscopic segregation of alloying elements [2-4].

As in any other multiphase material that combines two components with different mechanical properties, the mechanical contrast between the constituents can induce high local strains in the microstructure. These typically exceed the macroscopic strain because the soft phase (ferrite in a dual phase steel) has to compensate for the harder phase (martensite) to maintain strain compatibility [5]. Hence, any serious endeavour to understand the damage tolerance of these materials has to account for the complex interaction of the microstructural constituents and particularly their co-deformation behaviour. In a general sense, the principal stages of damage formation include the nucleation of microscopic voids in the microstructure, and their subsequent growth and coalescence with increasing strain until failure [6]. This process is conventionally studied post-mortem and largely manually, by analysis of samples obtained at different strains. Damage formation is, however, a dynamic phenomenon since the aforementioned stages occur and evolve depending on the microstructural environment, as well as the induced stress state. This is why in-situ techniques have become instrumental in unravelling dynamic processes as they allow researchers to observe the material deform directly and at the resolution of the chosen microscope [7-10]. In most cases, however, statistical relevance is sacrificed over spatial resolution by focussing only on a small area of the material to gain time- or strain-resolved information at high resolution. A further limitation is that data analysis of in-situ experiments is commonly done by hand so that larger datasets, including ones obtained post-mortem, remain difficult to analyse. To overcome these shortcomings of being limited by the field of view of a micrograph and the size of collected datasets at sufficient resolution, one has to adopt a much more powerful and efficient approach enabling 'high-throughput' ex- and in-situ analysis.

In the case of dual-phase steels, the pressure to achieve this, is clearly evident from the challenges still faced in unravelling damage. A vast number of studies exists from the macro- to the microscale, including a multitude of variations of the critical parameters, which have been identified [11-14]. However, so far it has proved difficult to reconcile all observations into comprehensive descriptions and models of damage in this important engineering material. We believe that the main reason lies in the heterogeneity of the microstructure, not only between samples, batches and alloys but also within each given volume of material [4, 15]. Where the high resolution studies of deformation mechanisms are intrinsically limited by their ability to study representative areas or volumes of the investigated material [5, 16, 17], the transfer and comparison of results between researchers become even more difficult. A new approach is therefore essential to propel our understanding of damage in complex advanced high strength steels. Such an approach will have to combine several essential aspects: (i) imaging of deformation-induced damage mechanisms at high resolution, (ii) efficient analysis to facilitate comprehensive comparisons of material state (e.g. strain level), (iii) integration of in-situ straining to observe damage evolution and (iv) quantitative analysis to the scale of a fully processed part (i.e. approaching mm² in sheet metal) to incorporate the inhomogeneous distribution of microstructural features [18] and variable stress/strain paths encountered during forming [1].

Here, we propose a promising new framework (Fig 1) to achieve just this, the analysis of large micrographs of the order of a mm² at high resolution, and therefore without the usual sacrifice of statistical relevance in the study of deformation-induced damage in highly heterogeneous microstructures. We employ panoramic imaging by scanning electron microscopy (SEM) at high resolution, both after deformation (ex-situ) and during straining (in-situ) inside the SEM. For this, we use a DP 800 dual phase steel as an example material that combines a high commercial popularity as a sheet metal in the automotive industry with great challenges in its characterisation and knowledge-driven improvement due to its intrinsic and scale-bridging heterogeneity from the micron to the millimetre scale. To enable automated detection and efficient quantitative and qualitative analysis of damage within the images recorded before, after and also during deformation, we use trained convolutional neural networks, known to be particularly apt at image recognition and analysis [19]. For almost a decade, the "ImageNet Challenge" [20] has been focussing on the optimisation of algorithms and techniques for image recognition and object detection and identification. In recent years, powerful architectures based on deep convolutional networks have started to dominate the field. The success of deep convolutional networks in image analysis has already found numerous applications, from the identification of cancerous anomalies in medicine [21, 22] to the detection of cracks in concrete [23].

In the following, we first describe the employed computational and experimental methods in detail before reporting our findings from bulk and in-situ deformation experiments. The results section is divided into the three main research questions concerning this topic, namely, the distinction between inclusions and deformation-induced damage, the definition of appropriate classes for deformation-induced damage, and, subsequently, the observed dependency of damage mechanisms on tensile strain in the dual-phase steel. The results are discussed and compared within both the fields of artificial intelligence and materials science and engineering.

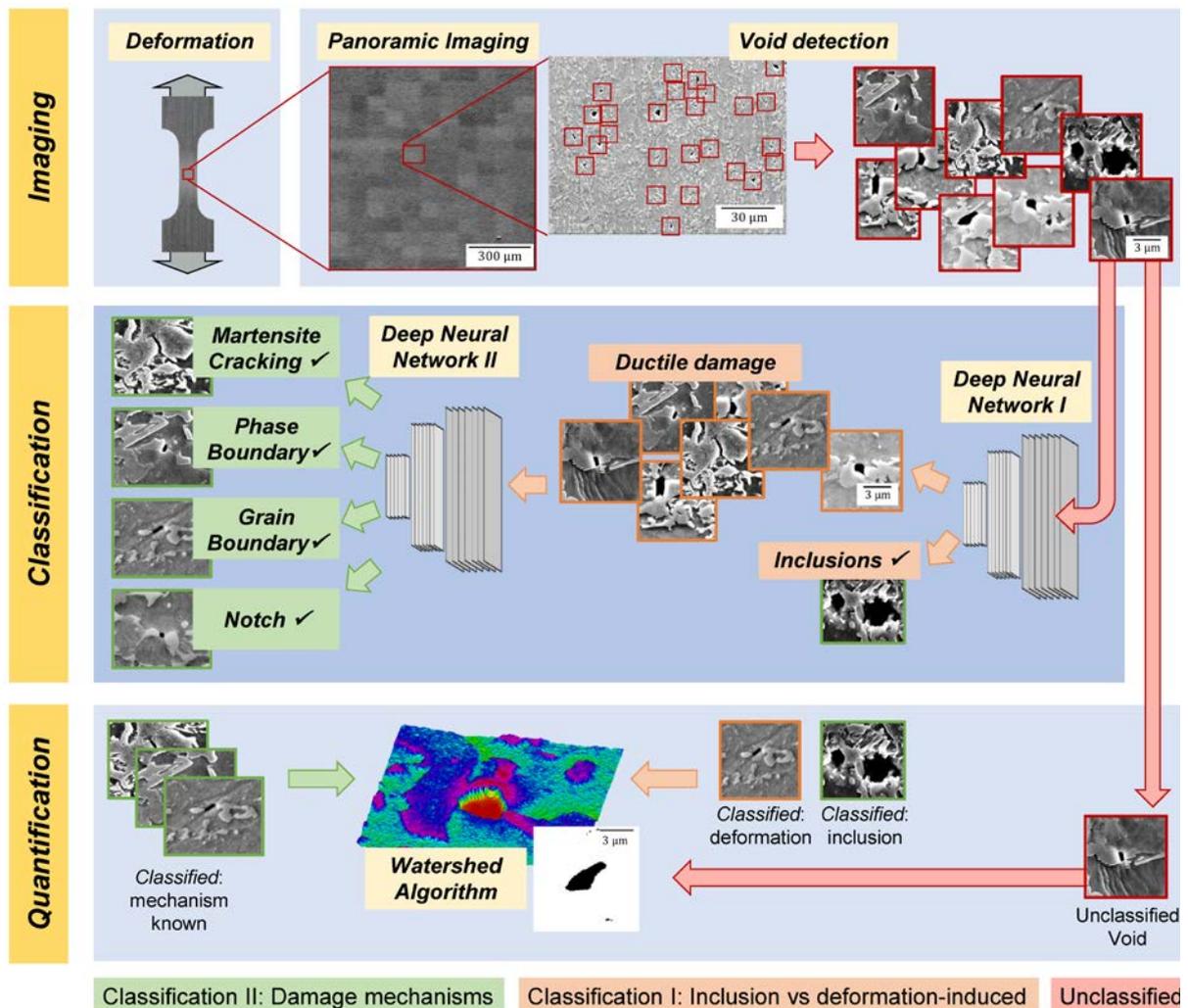

**Fig 1. New approach for high resolution statistical damage classification by artificial intelligence.** Based on panoramic imaging inside the SEM, the observed damage is identified, quantified and classified by artificial intelligence into four pre-selected categories (shown schematically).

# Methods

## Machine Learning and Neural Networks

Machine learning is a broad term which originally encompasses the "programming of a digital computer to behave in a way which, if done by humans or animals, would be described as involving the process of learning" [24]. In practice, this means using various algorithmic approaches to extract relationships from data and apply this knowledge to new samples. Popular machine learning algorithms include decision trees, support vector machines and in particular neural networks. On a fundamental level, neural networks mimic the basic working principle of a (human) brain (Fig 2): The basic building block is a neuron which receives multiple inputs, either from the "outside world" or other neurons, compute the weighted sum, which is then modified according to a transfer function before the output computed this way is passed to either the outside world or other neurons:

$$o_j = \sigma(\sum_i w_{ij} x_i) \tag{1}$$

where $o_j$ is the output of a specific neuron, $\sigma$ a transfer function, $w_{ij}$ a weight matrix describing the strength of the connection between input variables $x_i$ and the output.

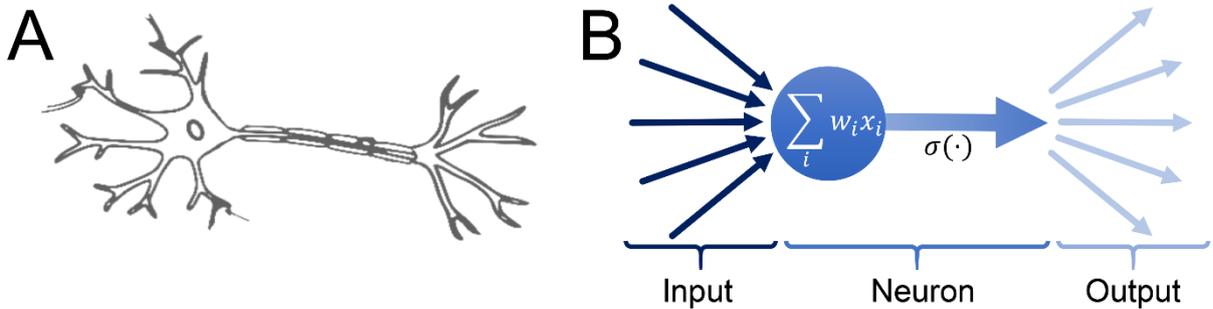

**Fig 2. Models of biological and artificial neurons.** (a) Simplified model of a biological neuron: The main body of the neuron receives input from many other neurons, transmits (or not) a signal which is then sent to many other neurons. (b) Model of an artificial neuron in a neural network. Input is received from other neurons or the outside world in the input layer, the computed output is transmitted to other neurons or the outside word for the output layer. (Image in (a) published with a CC0 license at https://pixabay.com/en/brain-neuron-nerves-cell-science-2022398/)

Using these neurons as building blocks, artificial neural networks are then built - in a simple picture - by combining many of these neurons as building blocks into layers. The first layer is called the "input layer" and receives the input from the outside world, in our case the micrograph images containing damage sites. The input is then passed through one or more intermediate or hidden layers to the output layer, which outputs the results to the outside world, in our case the researcher that receives the classification of the individual damage sites by the neural networks. As more and more powerful computer hardware has become available, neural networks can now consist of many such layers and are hence called "deep neural networks" (DNN). The exact shape of the weight-matrix allows to construct flexible and powerful DNN architectures which are ideally suited for different specialised tasks. In the case of images, so-called convolutional neural networks (CNN) have proven to be particularly apt in image recognition and analysis [19]. The name convolutional neural networks stems from the fact that the weights of the neurons are not free to take any value during network training but are constrained such that only few weights are available, which are shared across many neurons. These elements are arranged on diagonal bands in the weight matrix, e.g.

$$W = \begin{pmatrix} w_1 & \ldots & 0 \\ w_2 & w_1 & \ldots \\ 0 & w_2 & w_1 \end{pmatrix} \tag{2}$$

It can be shown mathematically that this is equivalent to the convolution of two distributions. Hence, CNNs can obtain the ability to act like filters in conventional image processing (e.g. edge detection, etc.). The main idea behind using CNNs is then that the various convolutional layers of the network can automatically learn the relevant features in the images associated with the content of the image, in our case a particular class of microstructural damage. The weight matrix of the final layer of the CNN network is fully populated and the individual weights are free to choose any value, i.e.

$$W = \begin{pmatrix} w_{11} & w_{12} & w_{13} \\ w_{21} & w_{22} & w_{23} \\ w_{31} & w_{32} & w_{33} \end{pmatrix} \qquad (3)$$

This so-called "fully connected" layer is used to translate the output of the CNN analysis into a probability associated with the identified damage class via a softmax function. The fundamental setup of the neural networks used in this analysis is then: input layer, several convolutional layers, a fully connected layer followed by softmax function.

The neural networks are trained using supervised training. At each iteration of the training process, a set of images is presented to the network for which the "true" output (label of damage sites) is known. Using the current state of network weights, the network calculates an output based on each image. Using an appropriate cost function, the difference between the calculated output and the desired (true) output is transformed into a penalty. The network weights are then iteratively adjusted via back-propagation, such that the discrepancy between the computed and desired output is minimal. One of the principal difficulties in training neural networks in general, and for this study in particular is access to sufficient high quality training data. Since no prior labelled training data was available, a dataset of - in total 4944 labelled damage sites - was obtained in the following way (Table 1): Potential damage sites where identified in panoramic electron microscope images using the same clustering approach as used in the later stages of the network training and application. The resulting candidate images were analysed and labelled manually by experts (material scientists). It is important to note that any bias the expert may have (e.g. showing a tendency to label images according to a given class or preferring one damage class to another) will translate into a bias of the final neural networks since the neural networks use these manually labelled images as ground truth. In order to remove such a potential bias, a blind study was performed among 19 experts including experienced researchers and students at undergraduate to post-graduate levels. Each expert was shown a set of 20 damage sites and the responses from the individual experts were cross-checked to ensure consistent assignment of labels to the damage classes. 80% of the dataset was used to train the neural networks, 20% was retained as an independent test sample to evaluate the performance.

**Table 1. Labelled training data over 4944 damage sites.**

| Damage class | Occurence (number of sites) |
|---|---|
| Inclusion | 572 |
| Grain boundary decohesion | 166 |
| Martensite cracking | 1582 |
| Notch | 874 |
| Phase boundary decohesion | 1750 |

The analysis setup is illustrated in Fig 1. As a first step, the grayscale high-resolution electron microscope panoramic image is analysed with respect to the grayscale values. As all damage sites considered in this analysis are characterised by a dark area in the micrograph, a suitable cut-off is chosen to identify the potential damage sites. Then a clustering algorithm (DBSCAN [25, 26], implemented in scikit-learn) is used to distinguish actual voids from artefacts like singular pixels below the cut-off value. The DBSCAN (Density-Based Spatial Clustering of Applications with Noise) algorithm takes a set of points as input, in our case the micrograph images. Fundamentally, algorithm group points that are close to each other (typically estimated by the Euclidean distance) are integrated into clusters. The algorithm is controlled by two parameters, the distance $\varepsilon$, which effectively describes how far the algorithm should consider points to be part of the current cluster, and the minimum number of points within a distance $\varepsilon$ that are needed to form a cluster. Points, which are not associated to clusters are considered as noise. A sample image of 250-by-250 pixels is then taken at each potential damage site from the panoramic micrograph. These candidate pictures are presented to a first deep convolutional neural network, which aims to identify whether the damage site in question is due to an inclusion. If the probability calculated by the neural network exceeds a pre-defined threshold ($p_1 > 0.7$), this damage site is classified as inclusion, otherwise, the image is cropped to 100-by-100 pixels and presented to a second deep convolutional neural network which is specifically trained to classify a damage site into martensite cracking (MC), notch (N), grain boundary decohesion (B) or phase boundary decohesion (PB). Again, if the calculated probability exceeds a given threshold ($p_2 > 0.7$), the damage site is classified accordingly, otherwise the picture is flagged for later manual analysis. Finally, the original panoramic microscope image is amended such that all identified damage sites are highlighted and labelled.

It was decided to split the total classification into two separate networks, for the sake of its use in application. As for many purposes like measurements of void area fractions of deformation-induced damage, only a separation into inclusion voids and deformation-induced voids is required, a first step should include this distinction exclusively. Then, in a second step, only the voids classified as deformation-induced should be processed further into classifying individual damage mechanisms. For the first network, a standard network (InceptionV3 [27]) was found to give best results (cf. Supplementary Materials), the architecture of the second CNN is given in Table 2, following Xu et al.[28]. In each case, ELU[29] was used as a transfer function. The input image is passed to the first convolutional layer as an appropriately sized input tensor of size (50,50,1). The much more complex architecture of the InceptionV3 network is described in detail in [27]. This difference in complexity can also be understood by regarding the used trainable parameters, 21,771,306 for InceptionV3, and 961,305 for the second network classifying the various damage mechanisms.

**Table 2. Architecture of the second deep neural network (CNN).** Descriptions of each layer type may be found in the supplementary materials.

| Layer type | # channels | Window Size | Stride |
|---|---|---|---|
| Convolution (input) | 192 | (5,5) | 1 |
| Convolution | 160 | (1,1) | 1 |
| Convolution | 96 | (1,1) | 1 |
| Max Pooling | | (3,3) | 2 |
| Dropout (p=0.5) | | | |
| Convolution | 192 | (5,5) | 1 |
| Convolution | 192 | (1,1) | 1 |
| Convolution | 192 | (1,1) | 1 |
| Max Pooling | | (3,3) | 2 |
| Dropout (p=0.5) | | | |
| Convolution | 192 | (3,3) | 1 |
| Convolution | 192 | (1,1) | 1 |
| Convolution | 10 | (1,1) | 1 |
| Average Pooling | | (8,8) | 1 |
| Dense + Softmax | | | |

Both networks were implemented in TensorFlow (v1.2.1) [30] using the Keras [31] API (v 2.1.5) with cuDNN (v 6.0.21) as backend. The clustering algorithms were implemented using scikit-learn [25].

The performance accuracy of the network was evaluated on 20% of the available data, which were retained as an independent test sample and not used in the training process. The accuracy is defined as the ratio of correctly identified images vs all available images in the sample:

$$a = \frac{\sum_{m=0}^{k-1} c_{m,m}}{\sum_{m=0}^{k-1}\sum_{n=0}^{k-1} c_{m,n}} \qquad (4)$$

where $c_{m,n}$ are the elements of the confusion matrix. The uncertainty of the accuracy was determined in the following way: The CNNs are initialised using pseudo-random numbers with variable seeds. This implies that in each training round the final network weights will be slightly different due to the numerical optimisations during the training. In this regard, the true positive rate (TPR, also: recall or sensitivity) was taken as the ratio of samples correctly identified as belonging to class m and the number of all samples belonging to this class, i.e.

$$TPR = \frac{c_{m,m}}{\sum_{n=0}^{k-1} c_{n,m}} \qquad (5)$$

and the positive predictive value (PPV, also: precision) as:

$$PPV = \frac{c_{m,m}}{\sum_{n=0}^{k-1} c_{m,n}} \qquad (6)$$

The CNNs were run on a Linux-based system, equipped with AMD FX-8340 CPU and 32 GB RAM. For the training a GeForce GTX 1070 GPU was employed. Since GPUs are non-deterministic in the implementation of their algorithms, slight variations are expected even if all other parameters are initialised to the same values. To estimate the size of the effect, each network was trained 10 times and the mean of the resulting spread in accuracy was taken as the point estimator, and the standard deviation as the uncertainty measure.

The training of a network took typically a few hours depending on the network architecture and more importantly on the number of training samples. Training on CPU would be possible in principle but training times are expected to be significantly higher by some orders of magnitude and have thus not been attempted. Training was executed for 50 epochs for the first network and 90 epochs for the second network. Both networks were trained using the Adam optimizer with a learning rate of 0.001 for the first network and a learning rate of 0.0005 for the second network, the remaining parameters were left at their default values, i.e. $\beta_1 = 0.9$, $\beta_2 = 0.999$, $\varepsilon = 10^{-8}$, and a learning rate decay of 0.0. See Supplementary Materials for accuracy and loss as a function of epoch as well as the respective confusion matrices for the test data set.

## Experimental Methods

The two dual phase steels used in this work are of commercial origin (ThyssenKrupp Steel Europe AG and ArcelorMittal SA, Luxembourg) and are marketed under the label "DP800", which warrants both a microstructure consisting of ferrite and martensite (dual-phase steel) and a tensile strength of at least 800 MPa. The microstructures differ in grain size and martensite morphology due to changes in processing and heat treatment. All tensile samples were cut out of the 1.5mm thick sheet metal by wire erosion, with the tensile axis being parallel to the rolling direction.

Metallographic preparation of all samples included mechanical grinding up to 4000 grit paper and mechanical polishing in diamond suspension in three polishing steps down to 1µm, followed by an additional final polishing step in a 0.25µm oxide suspension. All samples were lightly etched to ensure clear visibility of the ferrite and martensite constituents, while avoiding the build-up of a strong topography creating shadowing on the SEM images that might be detected as black areas by the void recognition system. For etching, a 1% Nital solution was used and the etching time was chosen to be 5 s to maintain the above-mentioned conditions.

The tensile samples were deformed by an electromechanical testing machine (DZM, in-house built) with an accuracy of ±0.17 MPa in the stress measurement. Strains in post-mortem experiments were measured as global strains over the deformed sample area based on the elongation between two markers in the undeformed parts of the sample. For in-situ tensile tests, a smaller deformation stage (Proxima 100, MicroMecha SAS, France [32]) was used. In-situ deformation strains, in addition to the initially applied pre-strain of the samples, were extracted from the deformation of the microstructure in the observed panoramic images by digital image correlation (DIC) and averaged over the observed area of 500 x 500 µm. All tensile specimens had a gauge length of 3.6 mm, width of 1.5 mm and thickness of 1.5mm (initial sheet thickness) to ensure comparability of strains and damage behaviour after necking between in-situ and ex-situ testing.

The panoramic images were acquired using scanning electron microscopes (LEO 1530 FEG-SEM, Carl Zeiss Microscopy GmbH, Germany & FEI Helios 600i, FEI Company, USA). All images were obtained using secondary electrons with a resolution of 32.5 nm/px (100 px horizontal field width in one image of 3072 px width). Image stitching for panoramic imaging was realised using MATLAB based on the VLFeat Matlab toolbox[33]. Damage recognition was realised by thresholding SEM grayscale images and usage of the DBSCAN algorithm before processing by the neural networks.

In the performed in-situ experiments, all samples were pre-strained to 14% global strain with reference to the initial gauge length of 3.6 mm. Additional strains were measured by averaging over a field of view in the panoramic imaging using a correlation between the individual panoramic images. The strains were averaged from the DIC measurements performed with GOM Correlate image correlation software (GOM GmbH, Germany). The so-obtained values therefore have to be regarded as average values, not over the complete sample, but rather limited to the investigated area by panoramic imaging.

# Results

## Separation of inclusions from deformation-induced damage

In understanding damage and deriving mechanism-based models of deformation in the dual phase microstructure of DP800 steel, one has to first distinguish between voids induced by the co-deformation of the ferrite and martensite phases and those induced around inclusions of foreign phases. Although the latter do not commonly cause a large amount of damage [34] , they can introduce significant errors in the quantification of damage if they are not distinguished from the deformation-induced damage. Conventionally, damage quantification by imaging techniques is achieved by a threshold analysis of greyscale image [35] to quantify the (darkest) areas within micrographs. Particularly where inclusions fall out of the surface during preparation, this may count them incorrectly towards deformation-induced damage. We therefore implemented a first neural network (CNN1 in Fig 1) to identify inclusions, both with and without a remaining foreign phase. Thereby, we enable a separate analysis of inclusion damage, which may strongly depend on manufacturer and batch quality, and deformation-induced damage, which is governed by the phase fractions and characteristics of the microstructure. Typically, these also depend on the manufacturer and can vary between batches depending on the alloy composition and the applied heat treatment.

The first step of our damage analysis is an automated acquisition of secondary electron micrographs with a resolution of 33 nm/px and a total field width after stitching to a single panoramic image of the order of 1 mm. In order to avoid a significant contribution of shadowing to the images yet be able to distinguish between ferrite and martensite phases, the commercial-grade DP 800 steel was subjected to light chemical etching beforehand. All voids within the panoramic micrograph were recognized via a thresholded image followed by a clustering algorithm (see Methods section for more detail on material, deformation experiments and algorithms employed in this work). Discrimination between those voids originating from inclusions in the steel matrix and others stemming from deformation-induced damage was then achieved by the first convolutional neural network presented with all clustered voids obtained from the previous step of void recognition. This initial step of damage analysis achieved an accuracy of $95 \pm 1$ %, PPV of $0.92 \pm 0.04$ and TPR of $0.85 \pm 0.07$ after training on a data set containing 572 manually labelled inclusions (out of a total of 4944 manually labelled damage and artefact sites).

The relevance of this approach is illustrated in Fig 3, where the number fraction of inclusion-induced voids is presented for a series of tensile samples deformed in small strain increments from a pre-strain of 12 % to a total strain of 24%. Particularly in the first one percent strain increment, the contribution of inclusions to the measured damage from a simple thresholding approach is considerable with over half of the measured damage sites pertaining to inclusion damage. With respect to area fraction, there is an even greater contribution of inclusion-related voids due to their average size of ~0.93 µm² compared with an average size of deformation-induced damage sites of ~0.43 µm². We believe that the ability to quickly and automatically remove inclusions from any further analysis of damage quantity or damage mechanisms is therefore of great value not only to achieve more accurate data but also to avoid superfluous experiments or valuable characterization time spent to quantify the density of inclusions before investigation of any new batch of material. This boosts not only the accuracy of studies regarding microscale damage mechanisms but is also important for quantifications of void area fractions in deformed samples without a focus on the underlying mechanisms, such as in the analysis of damage parameters for mesoscopic process simulations.

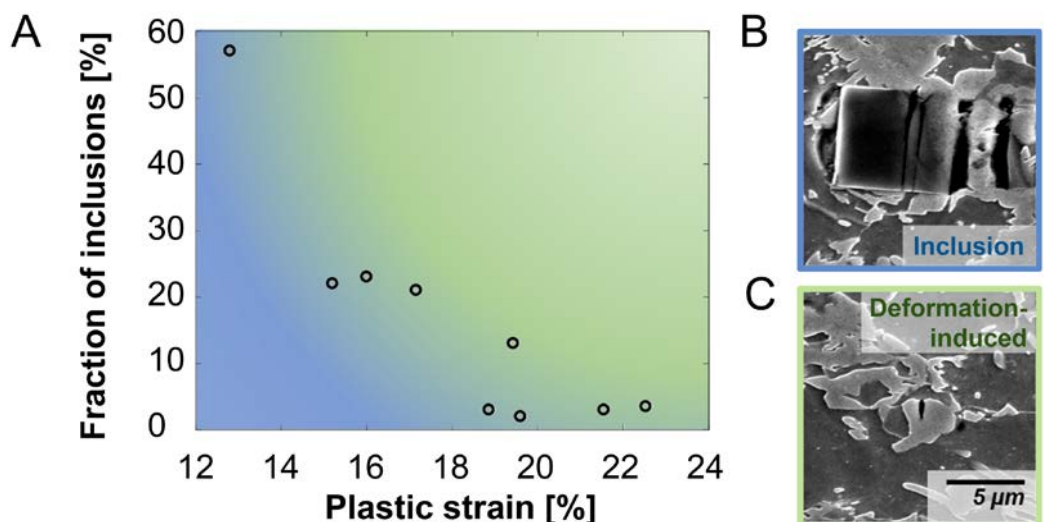

**Fig 3. Separation of inclusions and deformation-induced damage.** (a) Contribution of voids from inclusions to the total number of voids as a function of plastic strain in a tensile test (over an area of 0.63 mm²). Particularly at low strain levels the separation of inclusions and deformation-induced damage is essential. Micrographs illustrate the difference in size between (b) a typical inclusion and (c) a typical martensite crack induced as a result of deformation.

# Definition of classes of deformation-induced damage for statistical analysis

To generate a deeper understanding of the damage initiation during deformation, we started with the identification of fundamental damage mechanisms that are active in DP steel. According to a wide variety of investigations [36-38], those damage mechanisms can be narrowed down to ductile decohesion processes and fracture of brittle microstructure constituents. This results in the identification of the main mechanisms of void initiation as brittle martensite cracking and a ductile decohesion, dominated by ferrite plasticity, at either the martensite-ferrite interfaces or ferrite-ferrite grain boundaries. Note that the latter were very scarce in the present case due to their small number in the DP800 steel microstructure studied. Evolution of all types of voids, including those of martensite cracks once they have travelled across the affected martensite islands, was governed by plasticity of the surrounding ferrite grain(s).

However, our proposed approach in classifying damage sites in a statistically significant way is performed with a concept different from most previous studies – the most significant difference being the quantitative and qualitative analysis of large-area, high-resolution panoramic images. With, to our knowledge, only one notable exception[36], the studies on damage in DP steels currently available in the literature focus on relatively small areas or larger images recorded at low to medium resolution. They therefore deal with a limited number of evaluated damage sites and often these may be selected to correspond to particularly clear examples of the individual damage mechanisms considered. In contrast, our analysis aims to provide a statistical investigation of damage evolution and mechanisms and is therefore obligated to classify all occurring voids in a large field of view. Having approached our data with the above-mentioned set of damage categories, a large percentage (40 – 60% depending on the level of deformation) remains outside a reliable attribution to a certain damage mechanism.

We therefore propose to employ damage categories that not only capture the dominant deformation mechanism visible at the imaged stage but also reflect (i) any transition from brittle to ductile damage and (ii) the attainable resolution to resolve this transition. Furthermore, the categories need to be simple enough to allow reliable attribution to the individual categories independent of user experience or knowledge. This is important as deep learning involves training on labelled images and the success of the approach is therefore based on our ability to reproducibly assign distinct damage classification labels. A small study on 19 materials scientists ranging from undergraduate students to experienced researchers revealed that the application of the plethora of common damage classifications including cracking and decohesion of different boundaries is not an exact science. It is prone to great deviations between individuals where voids are regularly attached to both phases in a fine microstructure or damage has evolved significantly at advanced strain levels (note that the strain levels investigated here are actually at the low end between 12-24 % plastic strain compared with most other studies covering strains exceeding 50% in intervals of ~10% [10, 36]).

We therefore propose the following damage classes for a statistical and automated analysis of damage across high-resolution micrographs: (i) brittle martensite cracking, (ii) martensite-ferrite interface decohesion, and, (iii) a category labelled "notch" for short in which we collect those damage sites, which are too small to be reliably assigned to a distinct damage initiation mechanism. Mostly, these include round voids at thin martensite bridges, where one suspects the rupture of the martensite bridge due to the stress concentration to be the source of damage without the typical fracture faces that are usually clearly visible. Examples of each class are shown in Fig 1. We also explicitly labelled "evolved" damage sites that showed considerable evolution past the nucleation stage, particularly martensite cracks that had led to plastic flow in the adjoining ferrite grain(s) and therefore transitioned from pre-dominantly brittle to ductile deformation. We consciously included the category "notch" in addition to the classification of "evolved" damage in order to distinguish between those cases where it is the ductile evolution of damage that dominates the growth of the observed damage sites and those cases in which it is the lack of resolution at the sub-micron scale or strain increments that may prevent us from correctly distinguishing damage mechanisms in a freshly nucleated but diminutive void.

This leaves us with 4 clear-cut categories of damage (including the inclusion-induced voids discussed above) and the less distinct category of "evolved" damage. In order to achieve a clean training dataset for the neural networks, we therefore used only the former four categories for training of the neural networks. Clearly, the choice of training data and categories is therefore a critical point in which the main aim of the analysis is set for all following analyses using the same network. We chose to focus on the stage of nucleation of damage to achieve a clearer picture of the dominant damage nucleation mechanisms at different strain levels, leaving the analysis of evolving damage to the more appropriate method of in-situ experiments.

## Strain-dependence of deformation-induced damage

In order to identify the dominant damage mechanism(s) associated with a certain strain path, in this case uniaxial tension, the relevance of both the quality and quantity of results has to be ensured. In most DP steels, the martensite fraction is distributed heterogeneously across the material[1], mainly due to the formation of martensite bands in the rolled sheet metal. This directly necessitates the analysis of damage events across a large sample area to obtain a representative picture of the material's microstructure. This aspect is independent of the experimental method, whether it involves post-mortem or in-situ damage analysis.

Each of these methods has its strengths and weaknesses and we therefore employed both ex-situ and in-situ experiments in this study using damage quantification and classification via artificial intelligence to enable the analysis of suitably large areas. The results showed that the in-situ experiment yields significantly different damage characteristics due to evolution of damage at the free surface, as opposed to ex-situ experiments, where deformation occurs in the bulk and is revealed post-mortem. A direct application of damage classes defined using the morphologies observed after deformation in the bulk to in-situ experiments is therefore fraught with peril. In the following, we first present the ex-situ statistical analysis of damage at different strain levels followed by the results of the use of automated identification, classification and tracking of individual damage sites during their evolution in in-situ experiments.

To achieve the required size of analysed area capturing the variability in the microstructure we imaged an area of 900.000 µm² at a resolution of 32nm/px for each strain step reported below. Note that these areas were recorded in the sheet plane, but the dimensions would be equally applicable to cross-sections in rolling or transverse direction to capture the gradients in grain elongation, phase morphology and banding across a typical sheet thickness of 1 mm.

We present here the gathered data on detected damage sites from tensile tests to different strains with two purposes in mind: first, to present a statistical meaningful dataset, which covers strains from 12-24 % with smaller strain increments analysed to much greater statistical relevance than available elsewhere and, second, to illustrate the success of automating this laborious analysis by artificial intelligence. The latter step is crucial to make this kind of analysis possible on a regular basis in laboratories around the world. We can thereby finally overcome the challenge of comparison and transfer of data and derived models in this (and other) technologically important, yet immensely complex, class(es) of alloy(s).

The statistics reveal an exponential increase in the total number of damage sites with increasing tensile strain in the sample (Fig 4). This is commonly observed at larger strains but the expansion of this trend is confirmed here for smaller strains, which are at or below the detection level in studies with larger strain increments and reduced image size or resolution. The evolution of the total number of deformation-induced damage sites, i.e. excluding inclusions, as determined by human investigator and the neural networks is virtually the same within the scatter that remains even for the large areas analysed here (Fig 4).

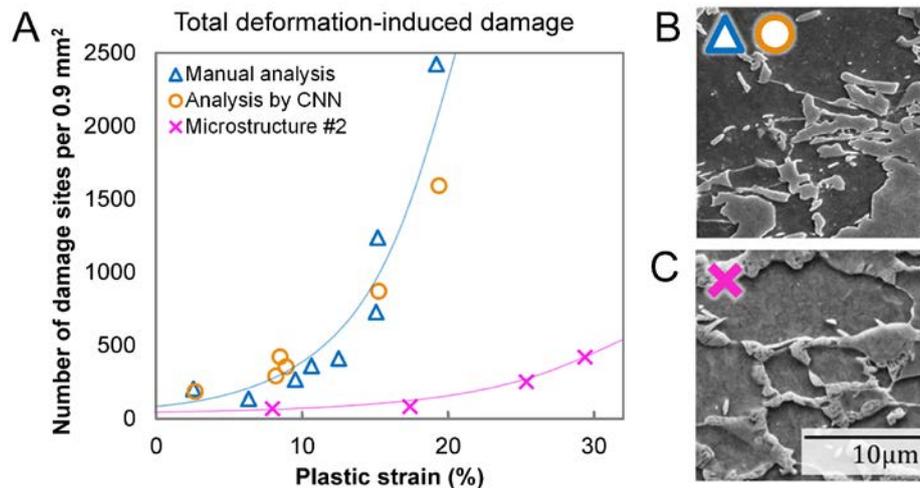

**Fig 4. Evolution of total damage as a function of uniaxial strain.** (a) Different panoramic micrographs were analysed manually (blue triangles) and by the CNNs (orange circles) both showing the same exponential increase in damage with strain. The drastic difference between microstructures of different DP800 grade steels is also shown for comparison (pink crosses). Micrographs of either microstructure are shown in (b) and (c).

The damage mechanisms classified as outlined above paint an interesting picture. The relative contribution of each damage class to the overall damage (Fig 5) shows a clear trend: the fraction of undeformed, newly nucleated martensite cracks decreases significantly with the strain level. This can be rationalised on the basis that martensite cracking at lower strains occurs predominantly in distinct regions of the microstructure, namely prior austenite grain boundaries [39]. These weak and therefore brittle locations tend to fracture at relatively low stresses, which causes martensite cracks to emerge predominantly in the early stages of deformation and cease afterwards. At larger strains, the ductile mechanisms of void nucleation (phase boundary decohesion) and void plasticity (evolving voids) play an increasingly dominant role in both the nucleation and evolution of damage voids. As a result, it may be essential to guide microstructure design concepts towards damage tolerance and even distribution rather than minimisation of damage, as suggested also by Tasan et al. [1].

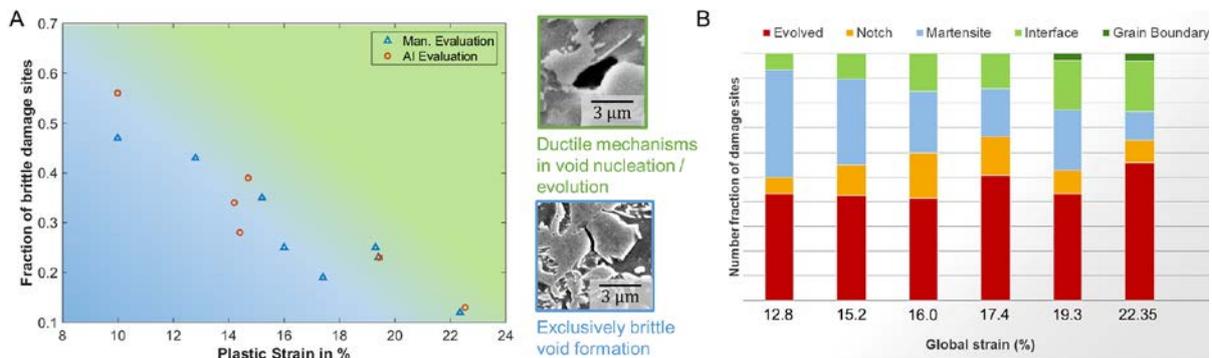

**Fig 5: Change of dominant damage mechanisms with strain.** The distinction between brittle and ductile damage (a) highlights the change in dominant mechanism with an only slowly increasing fraction of significantly evolved damage sites to the overall damage at increasing strains (b).

At this point, it is important to distinguish between the classification by a human investigator and the neural networks. As outlined above, the neural networks were trained using the categories inclusion, martensite cracking, phase boundary decohesion and notch. Strongly evolved damage sites, e.g. martensite cracks, which have already induced flow in the surrounding ferrite, were excluded in order to remove the bias of the training data, where different human investigators might not be confident that they can identify the underlying nucleation mechanism. As this distinction results in a training dataset with clearer dominance of each mechanism but cannot be based on distinct thresholds, the category of evolved damage cannot be implemented in a straightforward manner into an automated analysis by artificial intelligence. The classifications by a human investigator and the analytical algorithm consequently differ in some respects: first of all, not all sites are attributed to a mechanism by the human investigator, as a significant percentage is labelled as evolved damage sites. Most of these sites are, however, classified by the neural network as phase boundary decohesion sites due to their curved outline resulting from ductile flow, rather than brittle fracture across a straight path. While this might be a wrong classification in respect to the nucleating mechanism, it is consistent with ductile processes dominating evolution of voids at intermediate strains (Fig 5b) regardless of nucleation mechanism, which is the original definition of those evolved damage sites. The classification accuracy of the brittle martensite cracks is higher with $85 \pm 6$ % compared with the accuracy of identifying all classes ($80 \pm 6$ %). The division into voids that have an exclusively brittle origin and those that have nucleated or grown under the influence of ductile plastic flow in the nearest vicinity of the void is shown in Fig 4a.

To quantify the considerable differences in damage behaviour in different grades of commercial DP800 steel, a second microstructure containing a different martensite morphology was investigated. This material showed not only a higher ductility, but interestingly also a significantly lower number of nucleating damage sites during tensile deformation (Fig 4a). The observed behaviour underlines the need for large-area studies of each grade of investigated dual phase steel to achieve reliable predictions about their individual damage characteristics and exploit the full potential of the flexible processing conditions of these steels to achieve the most favourable, damage tolerant microstructures. These vital investigations can only be performed in a suitable tiescale with the help of automation.

# Discussion

In dual phase steels, investigation of a large area at high resolution is essential to provide comparability and transferability of results within this class of materials due to the scale of the microstructure and damage initiation at the sub-µm scale coupled with material heterogeneity up to the mm-scale. The results from our comparison of damage evolution in two different DP800 microstructures underlines this. The stark difference in damage densities with strain will make it necessary to statistically quantify and classify the damage mechanisms and their occurrence in every grade of DP-steel commercially available by combining high-resolution imaging at the scale of the damage mechanisms and large-scale observations to obtain sufficient statistics. To achieve an accurate and transferable results in characterising and, ultimately, predicting damage nucleation and evolution in any specific dual-phase microstructure, these time-consuming statistical analyses need to be conducted for each distinct material grade. Such an endeavour is in fact feasible via our automated approach as it is able to handle a high volume of damage sites in a drastically reduced amount of time compared to a human investigator.

Using a neural network to classify images into different damage categories leads us beyond the common threshold analysis of greyscale images. It allows us to distinguish not only the constituent ferrite and martensite phases and related pores from the surrounding matrix, but also distinguish other types of pores. In particular, image classification allows us to treat inclusions and the pores they cause where they may have fallen out during metallographic surface preparation separately from deformation-induced damage. This is of particular importance at small strains and where damage is quantified by means of an area fraction.

Owing to the large number of damage sites observed at high resolution, we could show for example that martensite cracking precedes what is often classified as phase boundary decohesion in later stages of damage evolution. Over a strain interval of 12 – 24 % plastic strain, we further demonstrated a pronounced change in governing damage mechanisms from brittle martensite fracture at the onset of damage to ductile damage initiation and evolution at higher strains. This is consistent with the observations from the literature, where martensite cracking is frequently reported as the dominant mechanism of damage initiation (e.g. Asik et al [40] ) and interface decohesion as the dominant mechanism at larger plastic strains (e.g. Hoefnagels et al. [36] ). As decribed by Azuma [41], the main cause for this transition from brittle to ductile damage can be found in the morphological evolution of voids. This is because such evolution often results in a convolution of different damage mechanisms; those which govern the stages of damage initiation, and those governing damage evolution. Naturally, this results in unclassified damage in the context of automated damage analysis focused on distinct nucleation mechanisms. This was also shown by Hoefnagels et al. [36] for the ductile growth of a brittle martensite fracture into the ferrite matrix. Although their large dataset recorded at medium resolution allowed only an indirect proof, it highlighted the need for high resolution images to support their hypothesis with more direct but, again, statistically less representative data. A comparison of the two datasets recorded at medium and high resolution [36] further highlights the importance of acknowledging the limit of resolution to which damage sites can be detected or categorized. In their quantification of damage in a biaxial tension experiment, the measured damage density increased by a factor of nearly 15 between the data measured at medium resolution over an area of 450,000 µm² and that identified over an area of 38,200 µm² at high resolution.

While statistical evaluation of deformed samples provides the important quantification of the number or area of damage sites and the dominance of each mechanism for a given strain level, it cannot deliver local information about the specific evolution of individual damage sites. To achieve this, additional in-situ tensile tests within an SEM were performed, completing the picture of damage statistics with time-resolved information at high resolution. The experiments were performed in such a way that a pre-strain of 14% was imposed prior to surface preparation for the in-situ observation, providing a pre-damaged state at the start of the in-situ test. This was necessary because the overall strain increment is usually limited by the formation of artefacts due to surface roughening from plastic flow at the open surface. The induced damage sites were then identified from the previously acquired panoramic image and classified in this state, which is equivalent to the post-mortem analysis presented above. Each site of interest can then be automatically tracked over subsequent in-situ deformation steps and pre-processed to achieve efficient analysis and interpretation as shown in Fig 6.

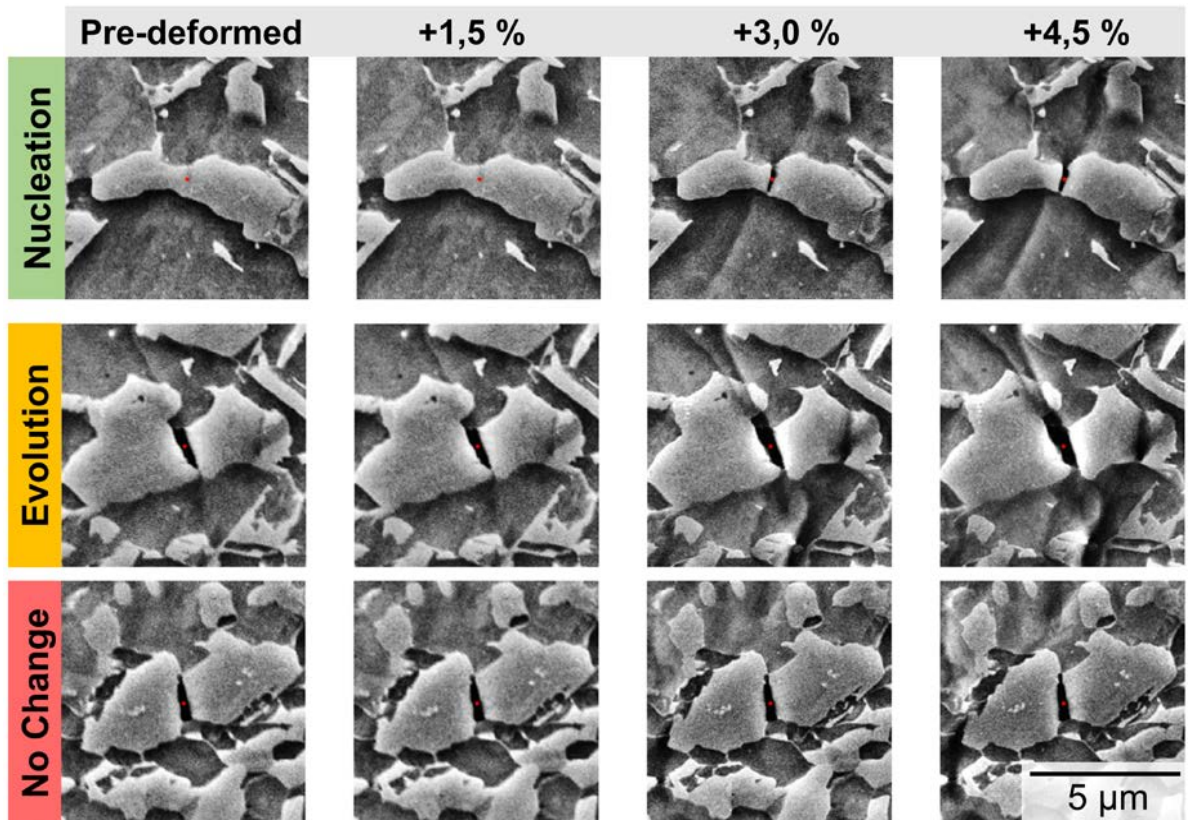

**Fig 6. Automatically detected and classified martensite cracks.** Crack development during in-situ experiments, showing (a) the nucleation stage and following crack propagation stage, (b) crack opening and evolution dominated by ductile behaviour in adjacent ferrite grains and (c) a similar martensite morphology as in (b) revealing no appreciable change over time at the same strain increments. The three-dimensional flow of the ferrite phase at the free surface in in-situ experiments leads to the formation of surface artefacts around the martensite islands in the form of trenches (indicated by white arrows in (b) and (c)), see also Fig 7.

Our in-situ tests show that the prevalence of nucleation of different types of damage sites is affected by the chosen method of investigation. For example, phase boundary decohesion, seen frequently in the post-mortem analysis of bulk samples, was barely encountered in the in-situ tests. Instead, plastic flow at the free surface caused "sink-in" of the ferrite matrix that led to the formation of deformation artefacts in the form of trenches (Fig 7). These artefacts, which are attributed to the altered stress state at the free surface of the deformed sample, were typically found to follow the martensite/ferrite interphase boundaries. The formation of surface artefacts during in-situ testing and damage analysis therefore poses a major difficulty in assessing the nucleation and evolution of phase boundary decohesion processes during plastic deformation as the dominant flow patterns in the bulk (see also Fig 7) and at the free surface clearly differ. Transfer of insights from in-situ experiments to bulk deformation, the dominant situation in metal forming due to the large volume to surface ratio, is therefore not possible with respect to ductile damage nucleation and evolution of brittle damage by flow. In this respect, in-situ experiments should be treated with great caution and used only to couple the plastic flow in the ductile phase observed in simulation and experiment where micromechanical crystal plasticity models are used to re-create the in-situ situation.

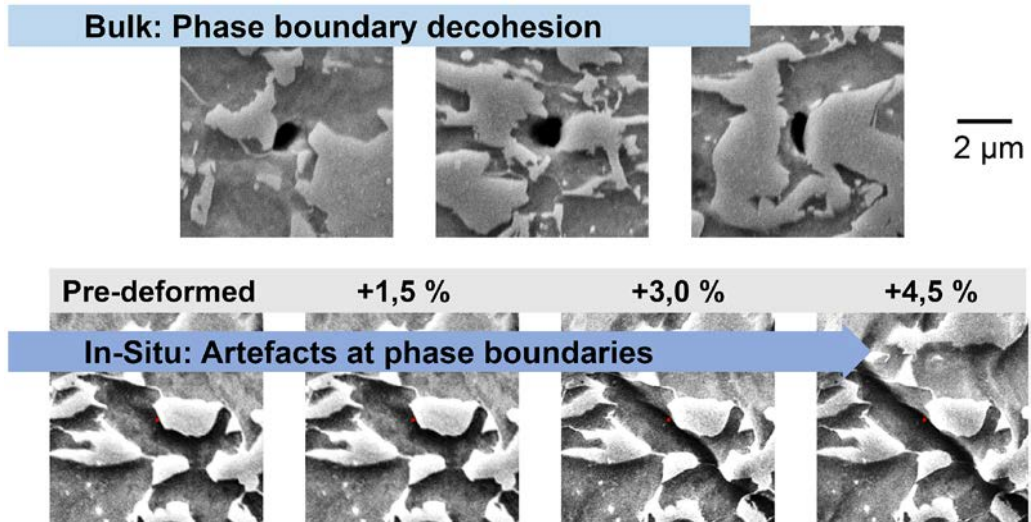

**Fig 7. Differences between bulk and in-situ experiments.** Plastic flow near phase boundaries between ferrite and martensite manifests very differently during deformation in the bulk and at the free surface in in-situ tests. In the bulk, it is dominated by phase boundary decohesion, while the surface is governed by artefacts due to flow normal to the free surface.

In contrast, the brittle mechanisms of martensite fracture appeared largely unaffected by the free surface in terms of their morphology. Further analysis of the nucleated martensite cracks during in-situ testing is therefore much more likely to shed a reliable light on the evolution of these damage sites with increasing strain level. This allowed us to not only track pre-existing damage site caused by martensite cracking in their evolution during the in-situ experiment, but also identify newly nucleated sites in subsequent deformation steps. These were then back-tracked to the pristine site (Fig 6a), giving us the opportunity to observe the onset of martensite cracking at numerous locations.

Across all sites of martensite cracking, we found that while some sites exhibited rapid growth as a result of plasticity in the ductile ferrite phase (Fig 5b), other sites showed virtually no change in comparison to the pre-strained state (Fig 5c). Intuitively, a first expectation about the evolution of an existing martensite crack would be that further growth occurs, assisted by plasticity in the ductile ferrite phase. However, this behaviour was not observed across all cracks identified in the pre-damaged state. For cracks that do not show any clear difference in morphology or surrounding microstructure, like those shown in Fig. 5 b) and c), the evolution process differs significantly from virtually no change in c) to a pronounced plasticity around the void and crack opening in b). Presumably, this is due to a change in the stress-state as the surface is cut free and half the surrounding martensite or ferrite structures are removed. The prevalence of this phenomenon is, however, important as it highlights how easily our view on damage evolution in in-situ experiments can be distorted, where only selected sites are reported, normally those that do show the evolution phenomena at the heart of most investigations.

The comparison of the results obtained in an in-situ test with our results on statistical damage analysis of material deformed in bulk (ex-situ) highlights the importance of large-area datasets and the observation of bulk deformation as the essential reference case, even where the experimental facilities for elegant in-situ tests are readily available. An ex-situ evaluation is unable to reveal the evolution of individual voids, however, it is essential to consider that the in-situ observation of void nucleation and evolution at the surface causes a substantial variation in void nucleation and evolution. This is widely acknowledged, but without the possibility of statistical analyses to reveal which damage mechanisms constitute the typical bulk case, crucial differences, like the altered void growth behaviour of martensite cracks at the free surface, are hard to quantify rigorously as an in-situ artefact caused by the free surface. Of course, the out-of-plane component of flow in in-situ experiments may still be considered in conjunction with crystal plasticity modelling, but so far, a truly quantitative evaluation is impeded by the experimental difficulties again encountered when combining high resolution and investigation of large areas [42, 43].

Our comparison of the statistical analysis of damage in bulk and that at the free surface reveals that in-situ tests are, however, representative in the context of brittle fracture in martensite. In this case, each individual martensite island present in the bulk or being part of the free surface may of course experience an altered stress state, leading to the absence of further growth of damage induced in the bulk during subsequent in-situ deformation. Still, the general morphology of martensite cracks remains the same and in-situ tests are therefore ideally suited to reveal the mechanisms of brittle damage nucleation. In-situ tracking of martensite cracks also included the ductile growth of a fraction of these originally brittle martensite cracks. This process portrays the transition to ductile mechanisms after brittle cracking. It was, in a similar way, observed in the material after bulk deformation as an increasing dominance of ductile mechanisms in void evolution (Fig. 5b). While the nucleation and evolution of damage by plastic deformation at the surface is difficult to interpret directly, it does reveal the assumed link between the nucleation and evolution stage: not only do ductile nucleation mechanisms become more dominant at larger global strains, but site-specific ductile mechanisms of void evolution take over at previously brittle nucleation sites of martensite cracking.

The neural networks developed in this study reduced the time required for micrograph analysis in terms of damage classification from days to minutes. While identifying and classifying a single damage site takes a human investigator about 20 s, leading to 5.5 hours per panoramic image with ~ 1000 damage sites, the use of trained artificial intelligence for this purpose reduces the time required to less than 1 minute (Fig 8). Depending on the required accuracy, the human investigator can instead focus on interpretation of the data and, where applicable, the few remaining unclassified cases, resulting in a total duration of a few minutes. In this context, we have also shown that a good performance can be attained based on a limited number of damage sites that can be acquired within a few hours. This was by no means pre-evident, as the underlying algorithms are commonly applied where millions of labelled, yet more variable and complex, images are available[44], e.g. from social media. Where mechanisms are of interest that occur rarely, i.e. a few times across an area of the order of 1 mm², but may be critical for material performance, the analysis by artificial intelligence is of particular help as it supports the researcher in the identification of atypical cases across large areas prohibitive to manual investigation. Such sites may then be represented in a new class of damage and fed back into the training dataset.

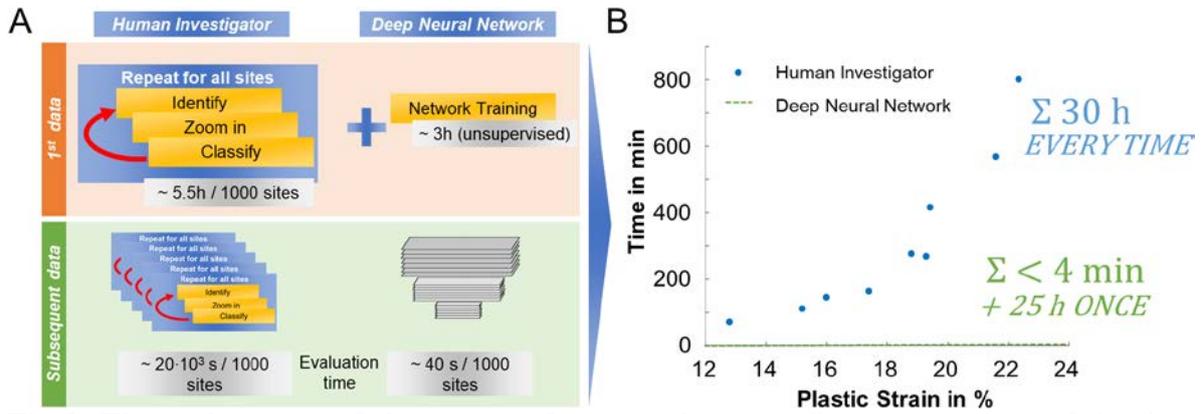

**Fig 8. Time advantage of the proposed approach.** (a) A direct comparison of the time invested (human investigator vs. artificial intelligence) in acquiring and processing initial and subsequent datasets of damage sites and (b) how the required time scales when evaluating strain dependent damage.

The use of automated recognition and analysis of damage sites by artificial intelligence in combination with in-situ testing is not an end in itself. Beyond the ability to evaluate representative datasets and save human supervision time on repeated analyses of similar microstructures, the approach enables studies that were previously inaccessible due to the constraints on human time; for example, the quantification of damage sequences in a statistical manner and coupling with high-throughput simulation to overcome the challenge of the missing 3D microstructure information where no suitable experimental methods [36, 45] are available. The method made available here brings together efforts in materials science and civil engineering. While the former community has used similar methods to those presented here in order to classify entire microstructures or their components [18, 46-48], the detection of damage has mostly been restricted to structural materials at the scale of buildings or bridges [49-51] and therefore investigated based on much lower resolution (photographic) images. This work links both worlds by considering damage, but, using electron microscopy, down to the microstructural level considered essential to the understanding, modelling and design of materials in high performance applications. We foresee many more applications of this approach in the future and hope that open availability of the developed neural networks and algorithms will encourage and enable others to apply these methods easily to their own investigations.

# Conclusions

In this study, we successfully implemented a high-resolution large-scale analysis with high throughput to provide damage quantification and classification in ex-situ and in-situ experiments. The following conclusions can be drawn:

- For complex, heterogeneous microstructures, such the ones encountered in dual-phase steels, the use of trained convolutional neural networks to classify a large number of damage sites at high resolution has proven extremely useful in gathering statistical information of representative damage mechanisms.

- For engineering purposes concerned with quantification of damage by means of an area fraction, it is prudent to split the automated classification work into two separate networks. The first network is to discriminate voids formed due to foreign inclusions from those induced by plastic deformation, and the second is to classify deformation-induced damage in conjunction with responsible damage mechanisms.
- A separate treatment of inclusion-induced and deformation-induced damage (as previously mentioned) provides a valuable improvement in accuracy for quantitative evaluation of damage statistics, especially at low deformation strains, where in the present case over half of the measured damage sites were affiliated with inclusion damage.
- The statistical viewpoint made possible by our approach unravels a more complete picture of the dominant deformation-induced damage mechanisms depending on the strain level. At incipient stages of deformation and damage initiation, brittle martensite cracking seemed to be the prevailing mechanism. An increasing amount of plastic deformation dominates at larger strains and led to rapid growth of voids and expansions of martensite cracks into the adjacent plastic ferrite phase.
- The damage behaviour of martensite cracking was reproduced in in-situ experiments since it was not affected by the surface artefacts generated upon straining. However, the morphology of the ductile damage sites were found to differ considerably from the bulk case. This highlights the need not to rely too heavily on in-situ experiments and interpret respective findings and their implications for bulk damage behaviour with great care.
- The present approach can, in principle, be transferred to similar image-based challenges in other complex microstructures at all scales. In the context of dual phase steels, a meaningful comparison of the manifold of microstructures subsumed under each industrial grade would be an exceedingly fruitful next step that now appears within reach. If successful, it would truly bring together the insights into the materials physics of deformation-induced damage, currently scattered across laboratories worldwide to enable more powerful knowledge-driven microstructure and process design for this important material class.
- Finally, any image recognition algorithm would have its limitation in real microstructures due to the complex interplay of individual damage formation mechanisms, rendering a major proportion of damage sites not clearly attributable to a single mechanism. Although we chose to focus on the mechanisms initiating damage in this work, we do note that in future work it should be feasible to further sub-categorize evolved damage based on sequential evolution stages associated with the onset of additional mechanisms.

The algorithms and networks along with a selection of image data are available online at https://git.rwth-aachen.de/Sandra.Korte.Kerzel/DeepDamage.git.

**Acknowledgments: -** (funding in submission system)